    \documentclass[british,english,showpacs,preprintnumbers,amsmath,amssymb,aps,notitlepage,twocolumn]{revtex4-1}
    \usepackage[T1]{fontenc}
    \usepackage{lmodern}
    \usepackage[latin9]{inputenc}
    \usepackage{comment}
    \usepackage{color}
    \usepackage{units}
    \usepackage{textcomp}
    \usepackage{amsmath}
    \usepackage{multirow}
    \usepackage{float}

    \makeatletter
    
    \providecommand{\tabularnewline}{\\}
    
    \usepackage{amsfonts}

    \makeatother
    
    \usepackage{babel}
    
\begin{document}
    
    %\title{\textcolor{black}{\normalsize{}Bound on quantum electrodynamics effects beyond $\alpha^4$ in the carbon atom with $\pm1$~cm$^{-1}$  precision}}
    \title{\textcolor{black}{ Computer-predicted ionization energy of carbon within $1$\,cm$^{-1}$ of the best experiment}}
    
    \author{Nike Dattani}
    \email{nike@hpqc.org}
    
    \affiliation{Harvard-Smithsonian Center for Astrophysics, Atomic and 
    Molecular Physics Division, 02138, Cambridge, MA, USA,}
    \affiliation{McMaster University, 606-8103, Hamilton, ON, Canada,}
%    \affiliation{Kyoto University, Department of Physics, 606-8103, Kyoto, Japan.}
    
    %\vspace{10mm}
    
    %\affiliation{Oxford University, Hertford College, OX1 3BW, Oxford, UK.}

    \author{Giovanni LiManni}
    \email{g.limanni@fkf.mpg.de}
    
    \selectlanguage{british}%
    
    \affiliation{Max Planck Institute for Solid State Systems, Department of Electronic Structure Theory, Suttgart, Germany.}

    \author{David Feller}
    \email{dfeller@owt.com }
    
    \selectlanguage{british}%
    
    \affiliation{Washington State University, Pullman, Washington
    99164-4630, USA,}
    \affiliation{University of Alabama, Tuscaloosa, Alabama
    35487-0336, USA,}

    \author{Jacek Koput}
    \email{koput@amu.edu.pl}
    
    \selectlanguage{british}%
    
    \affiliation{Adam Mickiewicz University, 61\textendash 614
    Poznan, Poland.}
    \selectlanguage{english}%
    
    %\vspace{10mm}
    
    \date{\today}
    \begin{abstract}
    We show that we can predict the first ionization energy of the carbon atom to within 0.872~cm$^{-1}$ of the experimental value. This is an improvement of more than a factor of 6.5 over the preceding best prediction in [Phys. Rev. A \textbf{81}, 022503], and opens the door to achieving sub-cm$^{-1}$ accuracy for \textit{ab initio} predictions in larger elements of the periodic table.% The sum of all error contributions in our calculations up to the fifth power of the fine structure constant is estimated to be at most 0.802~cm$^{-1}$, which yields our estimate of contributions in the sixth and higher powers to be: 0.994$~\pm~$0.802~cm$^{-1}$.  
    \end{abstract}
    \selectlanguage{british}%
    
    \maketitle
    \selectlanguage{english}%
    
    %\section{Introduction}
    \vspace{-20mm}
    %\vspace{10mm}
    
    In the last seven years, ionization energies (IEs) have been calculated
    with unprecedented precision for the Li atom \cite{*Wang2017,*Drake2018,Puchalski2010},
    Be atom \cite{Puchalski2013b} and B atom \cite{Puchalski2015}. Tight
    variational bounds for non-relativistic ground state energies assuming
    a clamped, point-sized nucleus have reached 49 digits in units of
    Hartree for He \cite{2006Schwartz}, 16 digits for Li \cite{Wang2017}, 12 digits for Be \cite{Puchalski2013}, and 11 digits for
    B \cite{Puchalski2015} (see Table \ref{tab:variational bounds}). Calculated
    IEs have been made in agreement with experiment to
    within $10^{-3}$~cm$^{-1}$ for $^{7}$Li, $10^{-1}$~cm$^{-1}$
    for $^{9}$Be and $1$~cm$^{-1}$ for $^{11}$B (see Table \ref{tab:excitationEnergies}). 
    
    %Boron agreement was 0.12cm-1. Not quite good enough to say "within 0.1cm-1"
    
    For the C atom, before this present study, no high-precision calculation had been reported to predict an IE to within $\sim1$~cm$^{-1}$
    agreement with experiment, but two new experimental papers have been published on this IE very recently \cite{Glab2018,Haris2017} as well as a new measurement of the electron affinity with more than an order of magnitude better precision than the best previous experiment \cite{2016Brestau}. Table \ref{tab:variational bounds} shows that the method used for the smaller atoms up to boron has not had success for carbon. With twice as many variationally optimizable parameters, one fewer digit was obtained for the B atom than for the Be atom, which also suggests that it would be very difficult to variationally optimize a fully explicitly correlated wavefunction ansatz for atoms and molecules coming from the rest of the periodic table. 
    
    The best known variational bound for the non-relativistic, clamped, point-sized nucleus (NR-CPN) ground state energy for
    C was calculated in 2015 using fixed-node diffusion Monte Carlo (FN-DMC)
    with the nodes of the electronic wavefunction fixed at the locations
    of a CISD/cc-pV5Z wavefunction, and the statistical uncertainty
    based on the stochastic fluctuations was $\pm20$~$\mu E_{\textrm{Hartree}}$ \cite{FN-DMC}. However, the IE for C predicted by FN-DMC was in discrepancy with experiment by more than 40\,cm$^{-1}$. In this
    paper, the approach we use to calculate the NR-CPN energy of the ground
    state of C is FCIQMC (full configuration interaction quantum Monte Carlo)
    with basis sets as large as aug-cc-pCV8Z. Table \ref{tab:variational bounds} shows
    that our NR-CPN energy is at least 76~$\mu E_{\rm{Hartree}}$ higher than the
    variational upper bound obtained from FN-DMC; but since in our approach, imperfections in the description of the wavefunction for the neutral atom are almost the same as in the cation, the individual errors almost completely cancel when
    calculating the energy difference. Therefore, with our approach we achieve agreement with experiment that is comparable within an order of magnitude to what has been seen with the explicitly correlated approach for atoms as big as (but not exceeding) boron. 
    
    %Table \ref{tab:variational bounds} shows
    %that the method used for the calculation has a crucial impact on the
    %precision obtainable: the best variational bound on the non-relativistic,
    %clamped, point-nucleus (NR-CPN), ground state energy of Li is believed
    %to be 8 orders of magnitude more accurate than that of the Be atom,
    %because analytic expressions for the multi-center integrals of Hylleraas
    %wavefunctions are only known for up to 3e$^{-}$, and it would be
    %very slow to calculate them numerically for 4e$^{-}$. For 4e$^{-}$
    %and greater, ECGs (\textbf{e}xplicitly \textbf{c}orrelated \textbf{G}aussian
    %wavefunction\textbf{s)} are used. ECGs mimic the theoretical shape
    %of the wavefunction less accurately (e.g. they are rounded rather
    %than cusp-like in the limit of vanishing electron-electron distance,
    %and in the infinite electron-electron distance limit they decay as
    %$\nicefrac{1}{r^{2}}$ rather than $\nicefrac{1}{r}$), so more parameters
    %are needed than when using Hylleraas wavefunctions. However, the integrals
    %of ECGs are known analytically, regardless of the number of electrons,
    %which is why it has been possible to do very high-precision calculations
    %on Be \cite{Puchalski2013b} and B \cite{Puchalski2015}. 
    
    %Table \ref{tab:variational bounds} demonstrates that with

    After adding relativistic and quantum electrodynamics (QED) corrections, and corrections to the clamped nucleus approximation,
    we obtained an IE for the ground state of C
    which is in only 0.872~cm$^{-1}$ disagreement with the best known experimental
    estimate. While this is not as impressive as the
    method of variationally optimizing parameters in an explicitly correlated
    wavefunction ansatz has proven to be for Li and Be, the disagreement with experiment 
    has the same order of magnitude as the latter approach for B (see
    Table \ref{tab:excitationEnergies}). We finally note that the approach
    used in this paper, of calculating FCIQMC on a basis set of non-explicitly
    correlated orbitals has successfully treated systems with far more
    electrons (transition metal atoms \cite{Thomas2015a}, diatomics \cite{Cleland2012}, multi-reference polyatomics such as ozone \cite{Powell2017}, 
    larger molecules such as butadiene \cite{Daday2012}, and even solid
    state systems \cite{Booth2012a}), so it is conceivable that the approach used in this paper may in the near future be able to determine (with
    fair accuracy) the IEs which at present remain experimentally elusive or poorly known according to NIST's atomic spectra database. These include arsenic (whose experimental IE has an uncertainty of $\pm2$\,\textrm{cm}$^{-1}$), Pm, Pa, Fm, Md, No, Sg, Bh and Hs (whose IEs are only known based on extrapolations of other experimental data and have uncertainties between $\pm$140\,cm$^{-1}$ and $\pm$4000\,cm$^{-1}$), Rf and Db (whose IEs are only known from theoretical calculations), and Mt, Ds, Rg, Cn, Nh, Fl, Mc, Lv, Ts, and Og (for which no IE is given in NIST's most recent databases).
    
    % http://physics.nist.gov/PhysRefData/Handbook/Tables/astatinetable5.htm doesn't give anything for At.
    
    \begin{center}
    \begin{table*}
    \caption{\label{tab:variational bounds}Upper bounds for
    total non-relativistic electronic energies. VO stands for variational optimization (parameters in a wavefunction ansatz are optimized to yield the lowest
    energy). Hylleraas-Log indicates the use of Hylleraas functions
    supplemented with auxiliary log functions, and ECG($M$) stands for  explicitly correlated Gaussian ansatz with $M$ optimizable parameters. Numbers in parentheses are estimated uncertainties \emph{within} the method used, so for FN-DMC does not include fixed-node error and for FCIQMC does not include basis set error. No numbers were obtained with basis set extrapolations.}

    %The uncertainties  given in parentheses do not include fixed-node error in FN-DMC or basis set incompleteness error in FCIQMC, so the lower ends of the confidence intervals may still be above the true variational limit. Knowledge of the size of these errors 
    
    \begin{tabular*}{1\textwidth}{@{\extracolsep{\fill}}ccr@{\extracolsep{0pt}.}lcccccc}
    \hline 
    \noalign{\vskip2mm}
     &  & \multicolumn{2}{c}{Total non-relativistic clamped point-nucleus (NR-CPN) energy [Hartree]} & Method/Ansatz type & Reference &  &  &  & \tabularnewline[2mm]
    \hline 
    \hline 
    \noalign{\vskip2mm}
    H & 1 & -0&5 & {Analytic \& Exact} & 1926 Schr\"{o}dinger &  &  &  & \tabularnewline
    He & 2 & -2&903 724 377 034 119 598 311 159 245 194 404 446 696 925 309 838 & VO/Hylleraas-Log & 2006 Schwartz & \cite{2006Schwartz} &  &  & \tabularnewline
    Li & 3 & -7&478 060 323 910 134 843 & VO/Hylleraas & 2017 Wang &    \cite{Wang2017}  &  &  & \tabularnewline
    Be & 4 & -14&667 356 494 9 & VO/ECG(4096) & 2013 Puchalski & \cite{Puchalski2013b} &  &  & \tabularnewline
    B & 5 & -24&653 867 537 & VO/ECG(8192) & 2015 Puchalski & \cite{Puchalski2015} &  &  & \tabularnewline
    C & 6 & -37&844 48(2) & {\footnotesize FN-DMC/CISD/cc-pV5Z} & 2015 Yang & \cite{FN-DMC} &  &  & \tabularnewline
    C & 6 & \textcolor{blue}{-37}&\textcolor{blue}{844~355~5(8)} & {\footnotesize FCIQMC/aug-cc-pCV8Z} & Present Work & - &  &  & \tabularnewline
    C & 6 & \textminus 37&843 333 & VO/ECG(1000) & 2013 Bubin & \cite{Bubin2013} &  &  & \tabularnewline[2mm]
    \hline 
    \end{tabular*}
    \end{table*}
    \par\end{center}
    
    \vspace{-6mm}
    
    \begin{center}
    \begin{table*}
    {\small \caption{\label{tab:excitationEnergies}The most precisely calculated ionization energies for the first six atoms, compared to the best known
    experimental measurements to date. The last column indicates that
    if aiming for the best precision, an experimental measurement is still
    the best way to obtain the energy for most atoms, but for Be, the
    energy has been obtained more precisely $\textit{in silico}$ than
    in any experiment to date. The value for carbon of \textcolor{red}{90~832.299}\,cm$^{-1}$ was calculated in the present work.}}
    
    \begin{tabular*}{1\textwidth}{@{\extracolsep{\fill}}lcr@{\extracolsep{0pt}.}lr@{\extracolsep{0pt}.}lr@{\extracolsep{0pt}.}lcr@{\extracolsep{0pt}.}lr@{\extracolsep{0pt}.}lc}
    \hline
    \noalign{\vskip2mm}
     & \multirow{2}{*}{Transition} & \multicolumn{2}{c}{Experiment } & \multicolumn{2}{c}{} & \multicolumn{2}{c}{Theory} &  & \multicolumn{2}{c}{Calc - Obs} & \multicolumn{2}{c}{$\lvert\frac{\text{Calc - Obs}}{\text{Uncertainty in obs}}\rvert$} & \multirow{2}{*}{More precise}\tabularnewline
     &  & \multicolumn{2}{c}{[cm$^{-1}$]} & \multicolumn{2}{c}{} & \multicolumn{2}{c}{[cm$^{-1}$]} &  & \multicolumn{2}{c}{[cm$^{-1}$]} & \multicolumn{2}{c}{} & \tabularnewline[2mm]
    \hline
    \hline
    \noalign{\vskip2mm}
    $^{1}$H & H$^{+}\left(1^{1}S\right)\leftarrow\text{H}\left(1^{2}S\right)$ & 109~678&771~732(23) & \multicolumn{2}{c}{\cite{Kramida2010}} & 109~678&771~743~07(10) & \footnote{This number is based on the data in \cite{PhysRevLett.95.163003} although it is not explicitly written anywhere there. Two of the authors of \cite{PhysRevLett.95.163003} have presented the number in Table III of \cite{RevModPhys.88.035009}. $^\textrm{b}$After the completion of this work two new values in disagreement with each other, $90\ 833.021(9)$ and $90\ 832.98(3)$, have been suggested and these are in even closer agreement with our result \cite{Haris2017,Glab2018}.} & 0&000~011 & 0&48 & Theory\tabularnewline
    $^{4}$He & He$^{+}\left(1^{2}S\right)\leftarrow\text{He}\left(1^{1}S\right)$ & 198~310&666~37(2) & \multicolumn{2}{c}{\cite{Kandula2011}} & 198~310&665~07(1) & \cite{Pachucki2017} & ~~~-0&001~3 & ~~~~~~65&00 & Theory\tabularnewline
    $^{3}$Li & Li$^{+}\left(1^{1}S\right)\leftarrow\text{Li}\left(2^{2}S\right)$ & 43~487&159~40(18) & \multicolumn{2}{c}{\cite{Bushaw2007}} & 43~487&159~7(7) & \cite{*Wang2017,*Drake2018} & ~~~-0&000~3 & ~~~~~~1&66 & Experiment\tabularnewline
    $^{9}$Be & Be$^{+}\left(2^{2}S\right)\leftarrow\text{Be}\left(2^{1}S\right)$ & 75~192&64(6) & \multicolumn{2}{c}{\cite{Beigang1983}} & 75~192&699(7) & \cite{Puchalski2013b} & ~~~0&059 & ~~~~~~0&98 & Theory\tabularnewline
    $^{11}$B & B$^{+}\left(2^{1}S\right)\leftarrow\text{B}\left(2^{2}P\right)$ & 66~928&036(22) & \multicolumn{2}{c}{\cite{Kramida2007}} & 66~927&91(21) & \cite{Puchalski2015} & ~~~-0&126 & ~~~~~~5&73 & Experiment\tabularnewline
    $^{12}$C & C$^{+}\left(2^{2}P\right)\leftarrow\text{C}\left(2^{3}P\right)$ & 90~833&171(15)$^\textrm{b}$ & \multicolumn{2}{c}{\cite{Chang1998}} & \textcolor{red}{90~832}&\textcolor{red}{299} & - & ~~~-0&872 & ~~~~~~~58&13 & Experiment\tabularnewline[2mm]
    \hline
    \end{tabular*}{\small \par}
    \end{table*}
    \par\end{center}

% 

%The uncertainty in the non-relativistic clamped-point-nucleus (NR-CPN) ionization energy comes from using two different basis set extrapolation formulas. Scalar relativistic effects are included via the X2C (exact 2-component) Hamiltonian, and a further correction is added via the Breit Hamiltonian and QED terms. Finite-mass corrections are added via the diagonal Born-Oppenheimer correction (DBOC). 

    \vspace{-20mm}
    
    %\vspace{-5mm}
    \vspace{-1.5mm}
    
    \section{Methodology}
    
    \vspace{-3mm}
    We begin with our main result in Table \ref{tab:Summary}, which shows that our computer-predicted ionization energy comes mainly from the NR-CPN Hamiltonian. This energy was calculated in four stages which we describe in the sub-sections below: (A) We developed larger core-valence (CV) basis sets than previously available for carbon, (B) we calculated the 1- and 2-electron integrals in these basis sets, (C) we solved the NR-CPN Schr\"{o}dinger equation at the FCI level in our finite-sized basis sets of two different sizes, and (D) we extrapolated the finite basis set results to estimate the energies at the \textbf{c}omplete \textbf{b}asis \textbf{s}et (CBS) limits. Finally, sub-section (E) describes how  we added the corrections due to special relativity, QED,  and due to the atom having an unclamped, zero-radius, nucleus.
    
    \vspace{-7mm}
    
    \begin{center}
    \begin{table}[H]
   % {
    \caption{\label{tab:Summary}{\small Summary of our main result. All energies are between the fine centre of gravity (fcog) of C$\left(^3P\right)$ and the fcog of C$^+\!\!\left(^2P\right)$ so the experimental spin-orbit lowering of 12.6725\,cm$^{-1}$ (calculated in our Supplemental Material, and based on measurements reported in \cite{Haris2017}) needs to be subtracted from all numbers to obtain the C$\left(^3P_0\right)\leftarrow ~ \textrm{C}^{+} \!\!\left( ^2 P_{\nicefrac{1}{2}} \right) $  energy. The experimental uncertainty is a 68\% confidence interval, meaning that there is a 32\% chance that the true energy is outside the range spanned by the uncertainty.}}

    \begin{tabular*}{1\columnwidth}{@{\extracolsep{\fill}}llr@{\extracolsep{0pt}.}lr@{\extracolsep{0pt}.}l}
    \hline
    \noalign{\vskip2mm}
    {Hamiltonian} &  & \multicolumn{2}{c}{{{}Ionization Energy}} & \multicolumn{2}{c}{\textcolor{black}{{}(Calc - Obs) }}\tabularnewline 
    & & \multicolumn{2}{c}{[\,cm$^{-1}$\,]} & \multicolumn{2}{c}{\textcolor{black}{{}[\,cm$^{-1}$\,]}}
    \tabularnewline[2mm]
    \hline
    \hline
    \noalign{\vskip2mm}
    NR-CPN       &  & 90~863&037 & \multicolumn{2}{c}{}\tabularnewline
    X2C          &  &    -30&023 & \multicolumn{2}{c}{}\tabularnewline
    Breit \& QED &  &     -0&48  & \multicolumn{2}{c}{}  \tabularnewline
    DBOC         &  &     -0&235 & \multicolumn{2}{c}{}\tabularnewline
    \noalign{\vskip2mm}
    \hline
    \noalign{\vskip2mm}
    Total (theory)  & Present & \textcolor{red} {90~832}&\textcolor{red}{299} &   \multicolumn{2}{c}{}\tabularnewline[2mm]
    \hline
    \noalign{\vskip2mm}
    Experiment   & 2017 \cite{Haris2017}& 90~833&021(9)  & \qquad  -0&722\tabularnewline
    Experiment   & 1998 \cite{Chang1998}& 90~833&171(15)  & \qquad  -0&872\tabularnewline %\tabularnewline[2mm]
    Experiment   & 1966 \cite{Johansson1966} & 90~833&122(100)  & \qquad  -0&823\tabularnewline[2mm]
    \hline
    \hline
    \noalign{\vskip2mm}
    Theory & 2010 \cite{Klopper2010}& 90~838&75 & \qquad 5&74\tabularnewline
    Theory & 2017 \cite{2017Feller}& 90~840&16 & \qquad 7&15\tabularnewline
    Theory & 2015 \cite{FN-DMC}& 90~786&66  & \qquad -46&35\tabularnewline[2mm]
    \hline
    \end{tabular*}
  %  }
    \end{table}
    \par\end{center}
    
    \vspace{-16mm}
    
    \subsection{{\scriptsize Optimization of `tight function' exponents for the aug-cc-pCV7Z and
    aug-cc-pCV8Z basis sets}}
    
    The largest orbital basis sets known for C prior to this work were the (aug-cc-pV$X$Z,
    $X=$7,8,9) sets used by Feller in 2016 \cite{Feller2016}. These basis sets
    did not contain `tight' exponent functions for capturing the effects
    of the correlation between the core $(1s^{2},2s^{2})$ electrons and
    the valence electrons $(2p^{2}$). The largest known basis set for carbon prior
    to this work including the CV (core-valence) correction was the aug-cc-pCV6Z \cite{Wilson1996} set. In this work we start by optimizing the `tight' exponents for
    the CV correction to Feller's 2016 aug-cc-pV7Z and aug-cc-pV8Z basis
    sets, yielding the first aug-cc-pCV7Z basis set for carbon, and the
    first aug-cc-pCV8Z basis set known for any element.
    
    The final aug-cc-pCV$X$Z basis sets have $X$ new tight functions of $s$-type, $X-1$ of $p$-type, $X-2$ of $d$-type, and so forth, up to the final $i$-type function for $X=7$ and the final $k$-type function for $X=8$. The $j^\textrm{th}$ exponent corresponding to a function of type $L$ is named $\gamma_{X,L,j}$, and is assumed to follow an `even-tempered' model: $\gamma_{X,L,j}=\alpha_{X,L,j}\beta_{X,L,j}^{j-1}$.
    
    In the non-linear optimization procedure to obtain $\alpha_{7,L,j}$ and $\beta_{7,L,j}$, the starting values were chosen to be the $\alpha_{6,L,j}$ and $\beta_{6,L,j}$ values that were already optimized in \cite{Wilson1996}. These were then treated as free parameters to minimize the difference between the frozen core and all-electron CISD energies of the carbon atom with all other exponent functions fixed. The $\tt{MOLPRO}$ program \cite{MOLPRO} was used to calculate the CISD energies, and the $\tt{L}$-$\tt{BFGS}$-$\tt{B}$ program of \cite{Zhu:1997:ALF:279232.279236} was used to optimize the free parameters. For $X=7$, the $s$-type functions were added first, then once they were optimized they were held fixed while the $p$-type functions were added and optimized. Then both the $s$- and $p$-type functions were held fixed while the $d$-type functions were added, and so on up to the single $i$-type function. The procedure for $X=8$ was the same, except the procedure continued to $k$-type functions, and the starting values came from the newly optimized $X=7$ case rather than the $X=6$ case from \cite{Wilson1996}. $\tt{MOLPRO}$ does not support $k$-functions, so to optimize the $k$-function we calculated the CISD energy at three points using $\tt{GAUSSIAN}$ \cite{g16} and estimated the value of $\alpha_{8,8,1}$ yielding the lowest energy by using a quadratic fit. 
    
    The tight exponents optimized in this work for aug-cc-pCV7Z and aug-cc-pCV8Z are presented in the Supplemental Material.
    
    %Table \ref{tab:tightFunctions}

    \vspace{-6mm}
    
    \subsection{{\scriptsize Calculation of 1- and 2-electron integrals including $k$- and $l$-
    functions}}
    
    %The calculation of the 1-and 2-electron integrals for the (aug)-cc-p(C)VXZ basis sets, that include up to k- and l- functions, 
    
    The calculation of the 1- and 2-electron integrals for (aug)-cc-p(C)V$X$Z
    basis sets with $X\ge7$ is not possible with most quantum chemistry
    packages, since  very few software packages support $k$-
    and $l$- functions, but for first row elements, $k$-functions appear in $X=7$ basis sets
    and $l$-functions appear when $X=8$. To calculate these integrals, we have used a locally modified version of {\tt MOLCAS} \cite{MOLCAS} in order to support larger basis sets. The 1- and 2-electron integrals for C and C$^+$ were evaluated in the basis of the optimized CASSCF(6,5) and CASSCF(5,5) orbitals respectively, with the five active orbitals being the 1$s$, 2$s$, 2$p_x$, 2$p_y$ and 2$p_z$ of the C atom/ion. This active space is the minimal active space including all electrons, that is able to provide balanced orbitals for the three degenerate states of the $^3P$ state of the C atom, or the $^2P$ state of the C$^+$ ion.
    
    %The MO-integrals have been evaluated in the basis of the optimized CASSCF(6,5) and CASSCF(5,5) MO orbitals, with the five active orbitals being the 1s, 2s, 2px, 2py and 2pz of the C atom. This active space is the minimal active space able to provide balanced orbitals for the three degenerate states of the 3P state of C atom

    %We have used ${\tt MOLCAS}$
    %to write the 1- and 2-electron integrals after a CAS(6,5)SCF  for C and CAS(5,5)SCF for C$^+$. In both cases, the 6 orbitals included were $1s,2s,2p_x,2p_y,2p_z$. The commercial version of ${\tt MOLCAS}$ had a
    %memory leak when printing the 56~GB file containing the 1- and 2-electron integrals for aug-cc-pCV8Z, so we needed to turn off the storage of density matrices which previously were calculated automatically in ${\tt MOLCAS}$. 
    
    \vspace{-5mm}
    
    \subsection{{\scriptsize Calculation of NR-CPN energies in finite basis sets without truncating the possible excitation levels (FCIQMC)}}
    
    \vspace{-3mm}
    
    A deterministic FCI (full configuration interaction) calculation for the 5e$^-$ C$^+$ ion in the aug-cc-pCV7Z basis set would require almost 55\,TB of RAM, and for the neutral atom would require more. Therefore we use FCIQMC for all NR-CPN calculations. %The wavefunction of the relevant system is expanded as a sum
    %of all possible Slater determinants given the basis set,
    %and the coefficients are determined by the number of walkers standing
    %on each determinant after a Monte Carlo sampling using the NR-CPN
    %Hamiltonian. 
    The method was introduced in \cite{Booth2009a}, and
    we use the initiator method first described in \cite{Cleland2010},
    and the semi-stochastic method as described in \cite{Blunt2015}.
    The calculations are performed using the developer version of the software ${\tt NECI}$ \cite{NECI}. 
    
    Within a given Hamiltonian (in this case the NR-CPN Hamiltonian) and
    basis set, there are three sources of error in the FCIQMC energy calculations: 
    \begin{enumerate}
    \item Trial wavefunction error ($\Delta E_{{\rm trial}}$), which approaches
    zero in the limit where the number of determinants used in the trial
    wavefunction approaches the number of determinants in the FCIQMC wavefunction; 
    \item Initiator error ($\Delta E_{{\rm initiator}}$), which approaches
    zero in the limit where the number of walkers $N_{{\rm walkers}}$
    gets sufficiently large; and 
    \item Stochastic error ($\Delta E_{{\rm stoch}}$), which for a given number
    of walkers and trial wavefunction determinants is estimated as the square root of the unbiased variance
    among different estimates $E_{i}$ of the energy from their mean $\bar{E}$
    after different numbers $N$ of Monte Carlo macro-iterations (determined using the Flyvbjerg-Petersen blocking analysis \cite{Flyvbjerg1989}) after the walkers
    have reached equilibrium: $\Delta E_{{\rm stochastic}}\approx\sqrt{\frac{\sum_{i=1}^{N}\left(E_{i}-\bar{E}\right)^{2}}{N-1}}=\mathcal{O}\left(\nicefrac{1}{\sqrt{N}}\right)$.
    \end{enumerate}
    
    Our goal was to obtain all energies to a precision of $\pm\epsilon$
    where $\epsilon\le1\mu E_{{\rm Hartree}}\approx0.2$~cm$^{-1}$ (within
    the basis sets used). To ensure that $\Delta E_{{\rm initiator}}$
    can be neglected, we used a sufficiently large value of $N_{{\rm walkers}}$
    for every energy calculation, so that the energy difference between
    using $N_{{\rm walkers}}$ and $\frac{1}{2}N_{{\rm walkers}}$ was
    smaller than 1 $\mu E_{\rm{Hartree}}$. Likewise, to ensure that $\Delta E_{{\rm trial}}$ can be neglected, we used a sufficiently large number of determinants in the trial wavefunction for every energy calculation, such that $\Delta E_{{\rm trial}}$ would also be smaller than 1 $\mu E_{\rm{Hartree}}$. We then ran every calculation for enough macro-iterations $N$ such that $\Delta E_{{\rm stoch}}$ was smaller than $\Delta E_{{\rm trial}}$ and $\Delta E_{{\rm initiator}}$. Further details are presented in the Supplemental Material, including tables which show that all three sources of error in our final numbers are not larger than we claim.
    
    \vspace{-6mm}
    
    \begin{center}
    \begin{table}[H]
     \caption{\label{tab:finalFCIQMC}{\small Final NR-CPN energies. The break-down of how
    these energies were obtained, and the Hartree-Fock energies that were
    used for the extrapolations are available in the Supplemental Material.
    Numbers within parentheses indicate uncertainties in the last digit(s)
    shown, and their determination is described in the Supplemental Material.}}
    
    \scriptsize
    \begin{tabular*}{1\columnwidth}{@{\extracolsep{\fill}}llr@{\extracolsep{0pt}.}lr@{\extracolsep{0pt}.}lr@{\extracolsep{0pt}.}l}
    \hline
    \noalign{\vskip2mm}
     & & \multicolumn{2}{c}{C$\left(^{3}P\right)$ } & \multicolumn{2}{c}{C$^{+}$$\left(^{2}P\right)$} & \multicolumn{2}{c}{$2^{3}P\rightarrow2^{2}P$~{}}\tabularnewline 
     & & \multicolumn{2}{c}{ $\left[E_{{\rm Hartree}}\right]$} &  \multicolumn{2}{c}{ $\left[E_{{\rm Hartree}}\right]$} &  \multicolumn{2}{c}{~[cm$^{-1}$]}
    \tabularnewline[2mm]
    \hline
    \hline
    \noalign{\vskip2mm}
    aug-cc-pCV7Z &  & -37&844~251~5(05) & -37&430~345~1(01) & 90~841&955(028)\tabularnewline
    aug-cc-pCV8Z &  & -37&844~355~5(08) & -37&430~412~5(05) & 90~849&987(054)\tabularnewline[2mm]
    \hline
    \noalign{\vskip2mm}
    Eq.\eqref{KutzelniggSolved}, $n=3.5$ &  & -37&844~528~6 & -37&430~523~6 & 90~863&604\tabularnewline
    Eq.\eqref{MartinSolved}, $n=4$ &  & -37&844~514~2 & -37&430~514~3 & 90~862&471\tabularnewline[2mm]
    \hline
    \noalign{\vskip2mm}
    Mean  &  & -37&844~521~4 & -37&430~519~0 & 90~863&037\tabularnewline[2mm] 
    \hline
    \end{tabular*}
    \end{table}
    \par\end{center}
    
    \vspace{-14mm}
    
    \subsection{{\scriptsize Extrapolations to the CBS (complete basis set) limit}}
    
    %An excellent review of the state-of-the-art in extrapolation choices is given on Page 5 of \cite{Feller2016}. The e 
    
    We use two different families of formulas to extrapolate the correlation energies (FCIQMC energies with the Hartree-Fock energies subtracted out) from $E_{X-1}$ and $E_{X}$ to $E_{\textrm{CBS}}$:
    
    %\begin{align}
    %E_{\textrm{corr,CBS}} & %=E_{\textrm{corr},X}-\frac{A}{X^{n}}~,\\
    %E_{\textrm{corr,CBS}} & %=E_{\textrm{corr},X}-\frac{A}{\left(X+\nice%frac{1}{2}\right)^n}~.
    %\end{align}
    
    \vspace{-5mm}
    
    \begin{align}
    \label{Kutzelnigg}E_{\textrm{CBS}} & =E_{X}-\frac{A}{X^{n}}~,\\
    E_{\textrm{CBS}} & =E_{X}-\frac{A}{\left(X+\nicefrac{1}{2}\right)^n}~.\label{Martin}
    \end{align}
    
    \noindent If we set $n=3$ in Eq.\,\eqref{Kutzelnigg}, we recover the formula originally proposed in \cite{Kutzelnigg1992}. If we set $n=4$ in Eq.\,\eqref{Martin}, we recover the formula originally proposed in \cite{Martin1996}.  If we have values for $E_{X}$ at two different $X$ values, we can eliminate $A$ in both cases, so Eq.\eqref{Kutzelnigg} leads to Eq.\eqref{KutzelniggSolved} and Eq.\eqref{Martin} leads to Eq.\eqref{MartinSolved}:
    
    \vspace{-2mm}
    
    \begin{align}
    \label{KutzelniggSolved}E_{\textrm{CBS}} & =\frac{X^nE_{X}-\left(X-1\right)^nE_{X-1}}{X^n-\left(X-1\right)^n}~,\\
    \label{MartinSolved}E_{\textrm{CBS}} & =E_{X}+\frac{\left(2X-1\right)^n\left(E_X-E_{X-1}\right)}{\left(1+2X\right)^n-\left(2X-1\right)^n}~.
    \end{align}
    
    As explained on page 5 of \cite{Feller2016}, extrapolations to the CBS limit using $n=3$ in Eq.\eqref{Kutzelnigg} tend to over-shoot the CBS limit. The value of $n=3.5$ was therefore used in \cite{Feller2016}, and we have used it in this present study. The values of $E_{\textrm{CBS}}$ obtained from using $n=3.5$ in Eq.\,\eqref{KutzelniggSolved} and $n=4$ in Eq.\eqref{MartinSolved}, for $X=8$ were added to the Hartree-Fock energies for $X=8$ and are presented in Table \ref{tab:finalFCIQMC}. The final NR-CPN energy was taken as the mean of both values obtained from extrapolating the correlation energy and adding it to the Hartree-Fock energy with $X=8$.%, and the uncertainty in this mean was estimated as the square root of the unbiased variance of the two values.
    
    \vspace{-7mm}
    
    \subsection{{\scriptsize Estimation of relativistic, QED, finite nuclear mass, and finite nuclear size corrections}}
    
    \vspace{-3mm}
    
    Scalar relativistic corrections were calculated by comparing the energies using the spin-free version of the 1e$^-$ X2C (exact 2-component) Hamiltonian \cite{Dyall1997,*Cheng2011}, to the energies of the NR-CPN Hamiltonian. %The spin-free 1e$^-$ version of the X2C Hamiltonian is equivalent to infinite-order DKH. 
    The integrals of our X2C Hamiltonian with ROHF orbitals were done in the $\tt{CFOUR}$ program, and were calculated at various levels of coupled cluster theory with the $\tt{MRCC}$ program \cite{MRCC,*Kallay2001}.
    
    Further scalar relativistic effects were included by adding the Breit and QED corrections
    (including the vacuum polarization and the self-energy terms that
    together comprise a Lamb-like shift) from the  state-averaged Dirac-Fock calculations done in \cite{Klopper2010}. The overall contribution from the Breit and QED correctons (combined) to the IE for C was -0.48\,cm$^{-1}$.
    
    Diagonal Born-Oppenheimer breakdown corrections (DBOC) to the clamped nucleus approximation \cite{KUTZELNIGG1997,*Gauss2006} were calculated using $\tt{CFOUR}$, with $\tt{MRCC}$ used for the coupled-cluster part. Our value of -0.235\,cm$^{-1}$ is about triple the value of -0.08\,cm$^{-1}$ estimated in \cite{Klopper2010}, due to including higher levels of correlation. 
    
    The basis set and correlation convergence of the X2C and DBOC corrections is shown in the Supplemental Material, along with our calculation of the finite nuclear size correction of 0.00543 cm$^{-1}$. The final corrections that contributed to our final computer-predicted ionization energy are presented in Table \ref{tab:Summary}.

\vspace{-6mm}
    
    \section{Conclusion}
    
\vspace{-4mm}
    
    Table \ref{tab:Summary} summarizes the various contributions to our value of the IE, and compares our final value to experiment and to three recent theoretical estimates. Our value is 0.872\,cm$^{-1}$ smaller than the best experimental value. %Since we included all known contributions up to the fifth power of the fine structure constant in the QED expansion of the energies, we estimate that the contributions of all terms of sixth power and beyond add up to (0.994$\pm$0.802)\,cm$^{-1}$.
    
    The best theoretical estimate of the IE before this work was in \cite{Klopper2010}, and was in disagreement with experiment by more than a factor of 6.5 more than our present result. We believe that this could have been due to any or all of three things: (1) approximations inherent to the F12 approach used for their NR-CPN energy, (2) the perturbative nature of their scalar relativistic corrections (i.e. using the mass velocity and Darwin terms, rather than the X2C Hamiltonian used in the present work) and (3) the CCSD approximation made in their DBOC correction  (as opposed to the CCSDTQ used in the present work which we have shown appears to be converged to the FCI limit).
    % IT OPENS THE DOOR FOR 6E- SYSTEMS LIKE LI2 [CITE]
    % ADD REFERENCES TO ORIGINAL FCIQMC AND INITIATOR AND EXCITED STATE PAPERS !!!
    
    \section{Acknowledgments}
    
    \vspace{-1mm}
    
    We wish to thank {\color{blue} Mariusz Puchalski, Krzysztof Pachucki, Robert Moszynski, Jacek Komasa, Gordon Drake, Michal Lesiuk, Michal Przybytek}, and {\color{blue} Wim Klopper} for helpful discussions, comments and suggestions. 
    
    \vspace{2mm}
    
    We also thank {\color{blue}Alexander Kramida} and {\color{blue}Kunari Haris} of NIST for information about their recent pre-print on a newer experimental ionization energy for carbon (mentioned in the footnote to Table  \ref{tab:excitationEnergies}), {\color{blue} Alexander Kramida} for information about the experimental ionization of hydrogen used in Table   \ref{tab:excitationEnergies}, and {\color{blue} Barry Taylor} and {\color{blue} Peter Mohr} of NIST for information about the theoretical ionization energy of hydrogen used in Table \ref{tab:excitationEnergies}.
    
    %Lan Cheng, Kirk Peterson, Willem Klopper, Alexander Lozovoi, and Ali Alavi 
    
    % Paul Ayers, Krzystof Szaliewicz, Sergiy Bubin, Cyrus Umrigar, Pablo Lopez-Rios, Richard Needs, Oleg Polyanski, Richard Dawes, Michael Bromley, David Ceperley, Peter Knowles, Sandeep Sharma, So Hirata, Piotr Peciuch, Jan Martin, Werner Kutzelnigg, Wenjian Liu, Grant Hill, Thom Dunning? Nortershauser? Csontos, David Tew, Willem Klopper, Zong-Chao Yan. NIST !!! Barry, Haris, Kramida! Sansonetti ? Pavenello, Babbush, Garnet Chan, Wiebe. Nikolaj Moll, Aspuru. George Booth, Robert Anderson. Frank Nesse, Jurgen Gauss. Nakatsuji. UBACHs !!!
    
    %\section{Dicsussion}
    
    %\clearpage
    %\newpage
    
    \selectlanguage{british}
%    \raggedright
%    \bibliographystyle{apsrev4-1}
%    \bibliography{bib.bib}
    \selectlanguage{english}

    \clearpage
    \newpage

    \end{document}